\documentclass[%
 aip,
 sd,%
 amsmath,amssymb,
reprint,%
]{revtex4-1}

\usepackage{graphicx}
\usepackage{dcolumn}
\usepackage{bm}
\usepackage{multirow}
\usepackage{color}

\begin{document}

\preprint{AIP/123-QED}

\title[]{Seed Layer Impact on Structural and Magnetic Properties of [Co/Ni] Multilayers with Perpendicular Magnetic Anisotropy}

\author{Enlong Liu}
 \email{Enlong.Liu@imec.be.}
 \affiliation{imec, Kapeldreef 75, Leuven 3001, Belgium.}
 \affiliation{Department of Electrical Engineering (ESAT), KU Leuven, Leuven 3001, Belgium.
} 
\author{J. Swerts}%
\affiliation{imec, Kapeldreef 75, Leuven 3001, Belgium.
}%
\author{T. Devolder}%
\affiliation{Center for Nanoscience and Nanotechnology, CNRS, Univ. Paris-Sud, Universit$\acute{e}$ Paris-Saclay, 91405 Orsay, France
}%
\author{S. Couet}%
\affiliation{imec, Kapeldreef 75, Leuven 3001, Belgium.
}%
\author{S. Mertens}%
\affiliation{imec, Kapeldreef 75, Leuven 3001, Belgium.
}%
\author{T. Lin}%
\affiliation{imec, Kapeldreef 75, Leuven 3001, Belgium.
}%
\author{V. Spampinato}%
\affiliation{imec, Kapeldreef 75, Leuven 3001, Belgium.
}%
\author{A. Franquet}%
\affiliation{imec, Kapeldreef 75, Leuven 3001, Belgium.
}%
\author{T. Conard}%
\affiliation{imec, Kapeldreef 75, Leuven 3001, Belgium.
}%
\author{S. Van Elshocht}%
\affiliation{imec, Kapeldreef 75, Leuven 3001, Belgium.
}%
\author{A. Furnemont}%
\affiliation{imec, Kapeldreef 75, Leuven 3001, Belgium.
}%
\author{J. De Boeck}
\affiliation{imec, Kapeldreef 75, Leuven 3001, Belgium.}
\affiliation{Department of Electrical Engineering (ESAT), KU Leuven, Leuven 3001, Belgium. 
}
\author{G. Kar}%
\affiliation{imec, Kapeldreef 75, Leuven 3001, Belgium.
}%

\date{\today}

\begin{abstract}
[Co/Ni] multilayers with perpendicular magnetic anisotropy (PMA) have been researched and applied in various spintronic applications. Typically the seed layer  material is studied to provide the desired face-centered cubic (\textit{fcc}) texture to the [Co/Ni] to obtain PMA. The integration of [Co/Ni] in back-end-of-line (BEOL) processes also requires the PMA to survive post-annealing. In this paper, the impact of NiCr, Pt, Ru, and Ta seed layers on the structural and magnetic properties of [Co(0.3 nm)/Ni(0.6 nm)] multilayers is investigated before and after annealing. The multilayers were deposited \textit{in-situ} on different seeds via physical vapor deposition at room temperature. The as-deposited [Co/Ni] films show the required \textit{fcc}(111) texture on all seeds, but PMA is only observed on Pt and Ru. In-plane magnetic anisotropy (IMA) is obtained on NiCr and Ta seeds, which is attributed to strain-induced PMA loss. PMA is maintained on all seeds after post-annealing up to 400$^{\circ}$C. The largest effective perpendicular anisotropy energy ($K_U^{\mathrm{eff}}\approx 2\times10^5$J/m$^3$) after annealing is achieved on NiCr seed. The evolution of PMA upon annealing cannot be explained by further crystallization during annealing or strain-induced PMA, nor can the observed magnetization loss and the increased damping after annealing. Here we identify the diffusion of the non-magnetic materials from the seed into [Co/Ni] as the major driver of the changes in the magnetic properties. By selecting the seed and post-annealing temperature, the [Co/Ni] can be tuned in a broad range for both PMA and damping. 
%
\end{abstract}

\maketitle

\section{Introduction}
Materials with perpendicular magnetic anisotropy (PMA) have recently received a lot of interest due to their use in spin-transfer-torque magnetic random access memory (STT-MRAM) and spin logic applications\cite{ITRS2015}. Magnetic tunnel junctions (MTJs) with PMA are required for further scaling of the critical device dimension (CD). The perpendicular MTJs (p-MTJ) enable STT-MRAM devices with longer data retention time and lower switching current at a smaller CD when compared to the MTJ’s with in-plane magnetic anisotropy (IMA)\cite{Driskill-Smith2011}. A typical p-MTJ stack comprises an MgO tunnel barrier sandwiched between a synthetic antiferromagnet (SAF) as fixed layer, and a magnetically soft layer as free layer. The material requirements for the fixed and free layer differ. Whereas the free layer is aimed to have a high PMA and low damping to ensure data retention and fast switching via STT, the SAF requires high PMA and has preferentially high damping to ensure that it remains fixed during STT writing and reading to avoid back hopping\cite{Devolder2016PRB}. As such, it is of high importance to control the PMA strength and damping in PMA materials. 

Co-based PMA multilayers [Co/X] (X = Pt, Pd) have received a lot of attention for their potential application in STT-MRAM, especially as SAF materials \cite{Yakushiji2010,Chatterjee2014,Chatterjee2015}. The PMA of these multilayers comes from the interface of [Co/X] in each bilayer repeat\cite{Johnson1996}. Besides, [Co/Ni] has been researched as alternative PMA material and has been employed in p-MTJ because of its high spin polarization and low Gilbert damping constant\cite{Mizukami2011,Tadisina2010,Tadisina2010a,Lytvynenko2015}. Also, [Co/Ni] has been incorporated in an ultrathin SAF\cite{Kar2014,Swerts2015, Tomczak2016}. Recently, the use of [Co/Ni] in the free layer material was proposed to enable free layers with high thermal stability needed at CD below 20nm\cite{Liu2016}. Next to STT-MRAM, [Co/Ni] has also been used as domain wall motion path in magnetic logic devices\cite{Bromberg2014}. Briefly speaking, [Co/Ni] multilayers are being considered for various applications in next generation spintronic devices. 
 
First-principle calculations predicted that [Co/Ni] in \textit{fcc}(111) texture possesses PMA. The maximum anisotropy is obtained when Co contains just 1 monolayer and Ni has 2 monolayers\cite{Daalderop1990,Daalderop1992}. Experimental studies proved that prediction\cite{den1992,Bloemen1992,Gottwald2012, Shioda2015} and reported on the PMA in [Co/Ni] for various sublayer thickness, repetition number and deposition conditions\cite{Beaujour2007,Gimbert2012,You2012,Haertinger2013,Akbulut2015}. To get [Co/Ni] with the correct crystallographic orientation and good texture quality, a careful seed selection is required. Various seed layers have been studied. as well as their impact on PMA and damping, including Cu\cite{Shaw2010,Wang2013}, Ti\cite{Song2013-2}, Au\cite{Kurt2010}, Pt\cite{You2012,Posth2009,Fukami2010,Gubbiotti2012,Ju2015}, Ru\cite{Sabino2014} and Ta\cite{Kato2011}. PMA change in [Co/Ni] is commonly observed and attributed to interdiffusion of the Co/Ni bilayers\cite{Wang2013,Cao2016}. However, we observed earlier that also the diffusion of the seed material can strongly impact the [Co/Ni] magnetization reversal, especially after annealing\cite{Liu2016}. A more in-depth study on the impact of the seed after annealing on the PMA and damping of [Co/Ni] is therefore required. Certainly the thermal robustness is of high importance for CMOS applications since the [Co/Ni] needs to be able to withstand temperatures up to 400$^{\circ}$C that are used in back-end-of-line (BEOL) processes. In this paper, we study four sub-5nm seed layers: Pt(3 nm), Ru(3 nm), Ta(2 nm) and Hf(1 nm)/NiCr(2 nm) and present their impact on both the structural and magnetic properties of as-deposited and annealed [Co/Ni]. We show that a good lattice match to promote \textit{fcc}(111) texture and to avoid [Co/Ni] interdiffusion is not the only parameter that determines the choice of seed layer. The diffusion of the seed material in the [Co/Ni] is identified as a key parameter dominating the PMA and damping of [Co/Ni] after annealing.

\section{Experimental Details}
The [Co/Ni] on various seed layers were deposited \textit{in-situ} at room temperature (RT) on thermally oxidized Si(100) substrates using physical vapor deposition system in a 300 mm Canon Anelva EC7800 cluster tool. Prior to seed layer deposition, 1nm TaN is deposited to ensure adhesion and to reflect the bottom electrode material that is used in device processing\cite{Liu2016}. The detailed stack structure  is Si/SiO$_2$/TaN(1.0)/seed layer/[Ni(0.6)/Co(0.3)]$_4$/Ni(0.6)/Co(0.6)/Ru(2.0)/Ta(2.0) (unit: nm). Ru/Ta on top serves as capping layer to protect [Co/Ni] from oxidation in air. The films were further annealed  at 300$^{\circ}$C for 30 min and 400$^{\circ}$C for 10 min in N$_2$ in a rapid thermal annealing (RTA) set-up.

The crystallinity of the [Co/Ni] films was studied via a $\theta$-2$\theta$ scan using the Cu $K\alpha$ wavelength of $\lambda=0.154$nm in a Bede MetrixL X-ray diffraction (XRD) set-up. The degree of texture is evaluated in the same tool by measuring full-width at half-maximum (FWHM) of the rocking curve in an $\omega$ scan with 2$\theta$ fixed at the \textit{fcc}(111) peak position of [Co/Ni]. Transmission electron microscopy (TEM) is used to identify the microstructure of multilayers. Time-of-Flight Secondary Ion Mass Spectrometry (ToF-SIMS) is used to study diffusion of the seed material in the [Co/Ni]. The measurements were conducted in a TOFSIMS IV from ION-TOF GmbH with dual beam configuration in interlaced mode, where O$^{2+}$ and Bi$^{3+}$ are used for sputtering and analysis, respectively. X-ray Photoelectron Spectroscopy (XPS) is used to quantify the diffusion amount of the seed layers. The measurements were carried out in angle resolved mode using a Theta300 system from ThermoInstruments. 16 spectra were recorded at an exit angle between 22$^{\circ}$ and 78$^{\circ}$ as measured from the normal of the sample. The measurements were performed using a monochromatized Al $K\alpha$ X-ray source (1486.6 eV) and a spot size of 400$\mu$m. Because of the surface sensitivity of the XPS measurement (depth sensitivity is $\pm$5 nm), the XPS analysis was carried out on [Co/Ni] films without Ru/Ta cap. 

A Microsense vibrating samples magnetometer (VSM) is used to characterize the magnetization hysteresis loops and to determine the saturation magnetization ($M_s$). The effective perpendicular anisotropy field ($\mu_0H_k^{\mathrm{eff}}$) and the Gilbert damping constant ($\alpha$) in [Co/Ni] was measured via vector-network-analyzer ferromagnetic resonance (VNA-FMR)\cite{Bilzer2007} and corresponding analysis\cite{Devolder2013}. The effective perpendicular anisotropy energy ($K_U^{\mathrm{eff}}$) is calculated by the equation $K_U^{\mathrm{eff}} = \mu_0H_k^{\mathrm{eff}}\times M_s/2$ in the unit of J/m$^3$.

\section{Results and discussion}
\subsection{Seed layer impact on structural properties of [Co/Ni]}
\begin{figure*}
\centering
\includegraphics[scale=0.25]{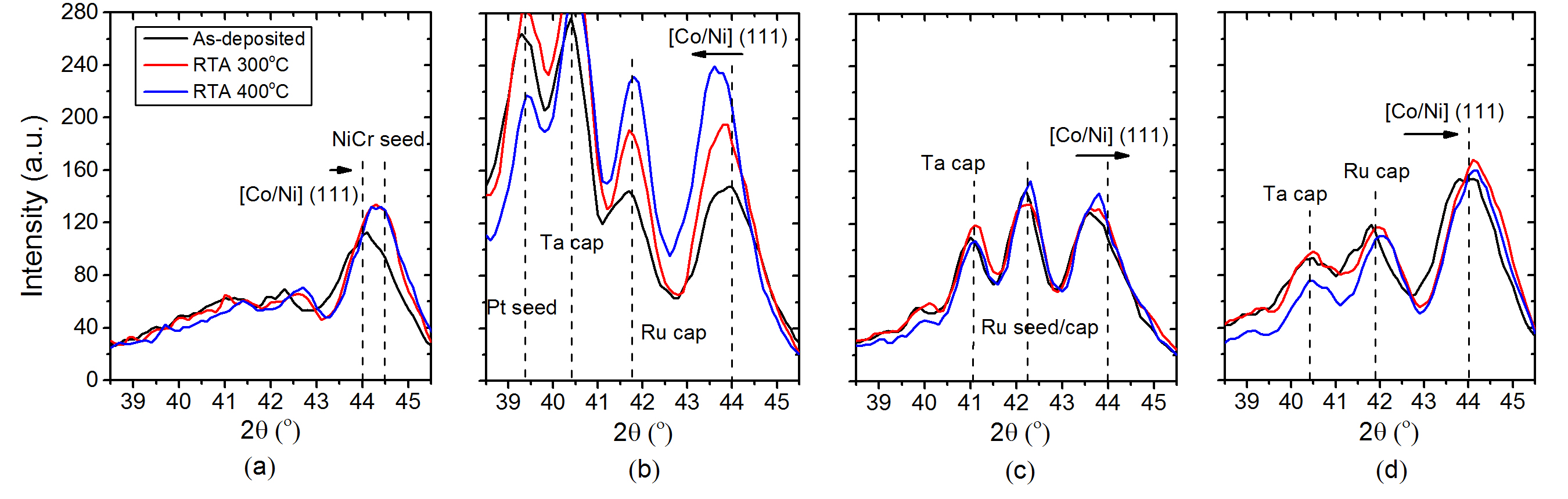}
\caption{$\theta$-2$\theta$ XRD patterns of [Co/Ni] on four seed layers:  (a) NiCr, (b) Pt, (c) Ru, (d) Ta, as-deposited, after 300$^{\circ}$C/30 min annealing and after 400$^{\circ}$C/10 min annealing, respectively. The dashed lines reflect the peak position of the NiCr seed, Pt seed, Ru seed/cap, [Co/Ni] and Ta cap. The arrows indicate the peak shift after annealing.}
\label{F1}
\end{figure*}

To obtain large PMA in [Co/Ni] systems, highly textured, smooth \textit{fcc}(111) films are required\cite{Gimbert2012}. Fig.1 shows the $\theta$-2$\theta$ XRD patterns of [Co/Ni] on different seed layers as-deposited and after annealing. The \textit{fcc}(111) peaks of bulk Co and Ni are located at 44.1$^{\circ}$ and 43.9$^{\circ}$, respectively\cite{davis2000nickel}. The presence of a peak around 2$\theta$ = 44$^{\circ}$ confirms the \textit{fcc}(111) texture in as-deposited [Co/Ni] films on all seed layers\cite{Rafaja2000}. After annealing, the peak intensity increases, in particular for the [Co/Ni] sample on a Pt seed. Additionally, a shift in the peak position is observed for all seeds. The black arrows indicate the shift direction in Fig.\ref{F1}(a)-(d). For [Co/Ni] on NiCr and Ta seed, the diffraction peak of [Co/Ni] shifts towards larger peak position of bulk Co and Ni, meaning that [Co/Ni] films on NiCr and Ta possess tensile stress after annealing. In contrast, compressive stress is induced in [Co/Ni] on Pt seed, since the peak position moves towards the Pt(111) peak due to lattice matching. On Ru seed, the peak nearly does not shift after annealing.    

\begin{figure}
\centering
\includegraphics[scale=0.27]{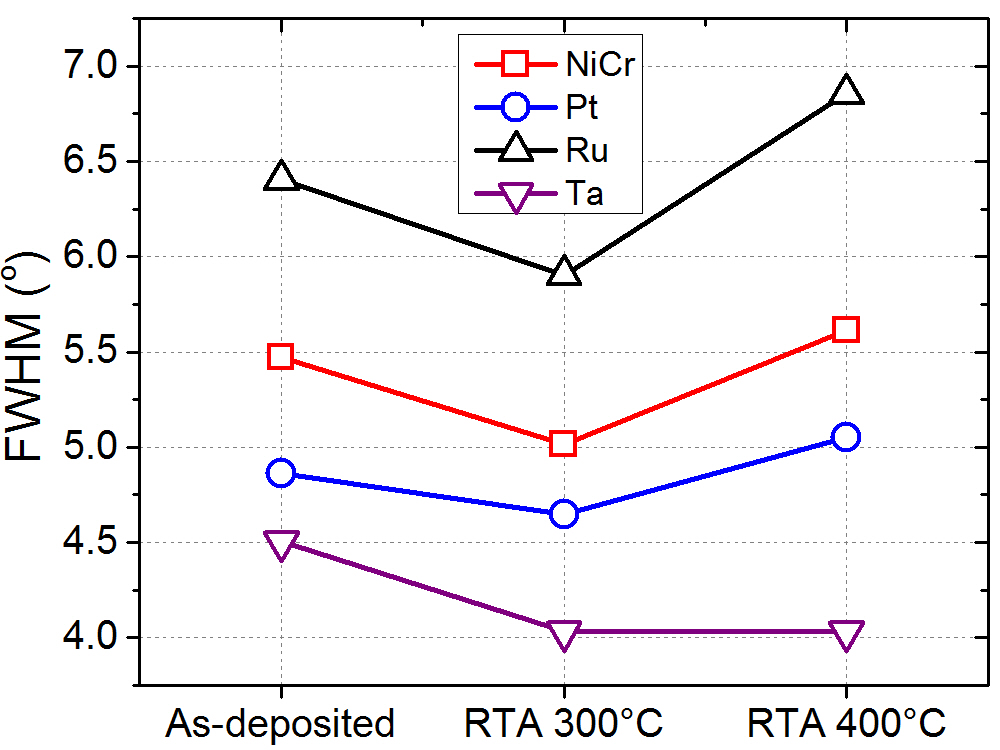}
\caption{FWHM of $\omega$ scan of as-deposited and post-annealed [Co/Ni] on different seed layers.  }
\label{F2}
\end{figure}

The quality of \textit{fcc}(111) texture of [Co/Ni] is further examined via rocking curves. The results of FWHM of the rocking curves are given in Fig.\ref{F2}, as well as the influence of post-annealing. The larger FWHM observed in [Co/Ni] on Ru seed suggests a lower degree of texture, which is in agreement with the $\theta$-2$\theta$ pattern where [Co/Ni] on Ru seed shows a peak with lower intensity. Post-annealing at 300$^{\circ}$C leads to further crystallization and enhanced texturing, as indicated by the decreased FWHM for [Co/Ni] on all seed layers. However, FWHM increases after 400$^{\circ}$C annealing, which may be attributed to the intermixing of Co and Ni and hence the degradation in crystal quality.

\begin{figure*}
\centering
\includegraphics[scale=0.4]{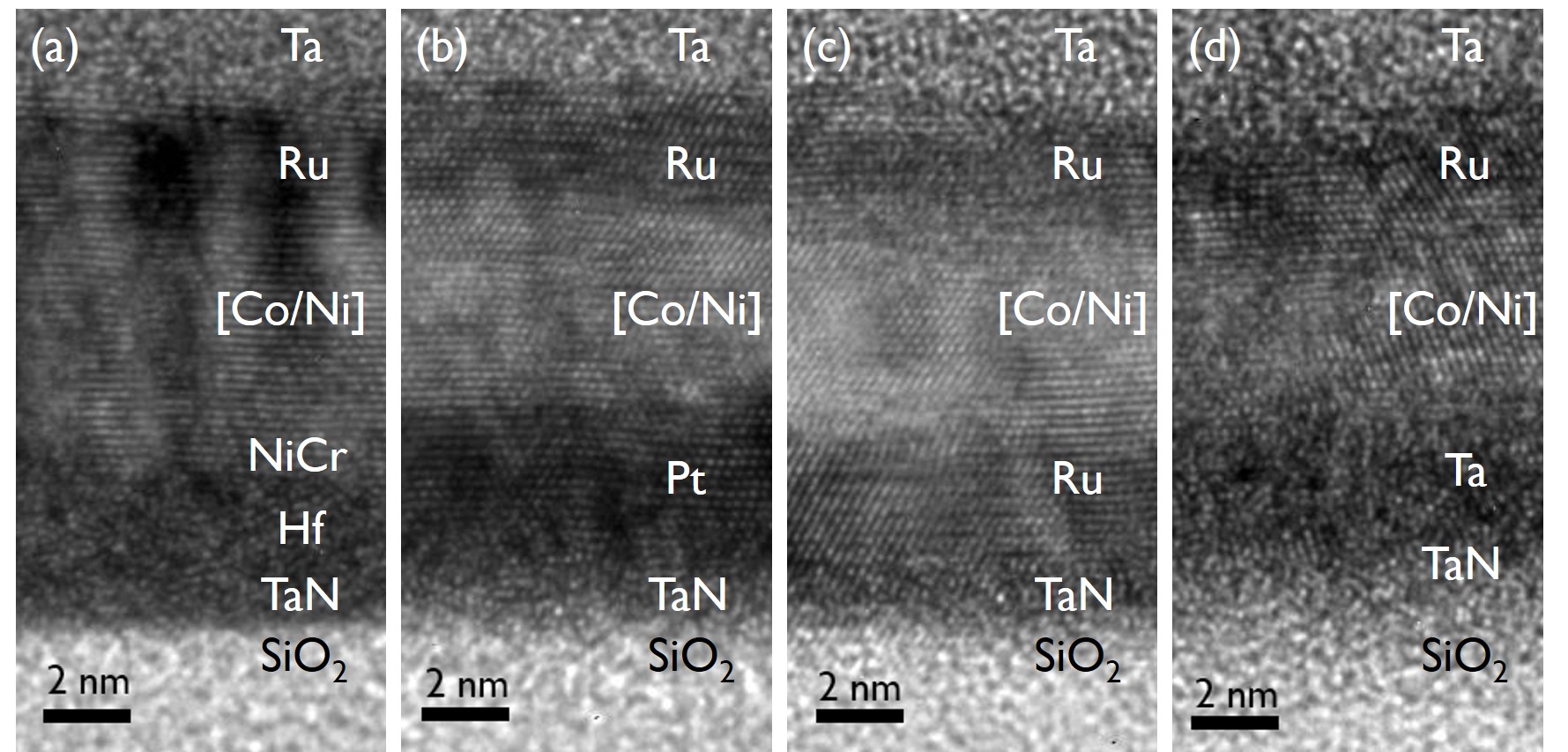}
\caption{TEM images of the microstructure of [Co/Ni] deposited on different seed layers after 300$^{\circ}$C annealing for 30 min: (a) Hf/NiCr, (b) Pt, (c) Ru and (d) Ta .}
\label{F3}
\end{figure*}

Fig.\ref{F3} shows the microstructure of [Co/Ni] deposited on different seed layers imaged by TEM after 300$^{\circ}$C annealing. In all [Co/Ni] samples the grains extend from seed to cap. The interface between [Co/Ni] and seed layer is clear and smooth in the samples with Pt and Ru seed layer in Fig.\ref{F3}(b) and (c), respectively. For [Co/Ni] on NiCr, however, the interface between the multilayers and the seed layer cannot be distinguished (Fig.\ref{F3}(a)), but both show clearly crystalline and texture matched. For the [Co/Ni] on Ta seed in Fig.\ref{F3}(d), the interface is quite rough and a nanocrystalline structure at the interface between Ta and [Co/Ni] can be spotted, which may indicate intermixing. 

\subsection{Seed layer impact on magnetic properties of [Co/Ni]}
\begin{figure}
\centering
\includegraphics[scale=0.3]{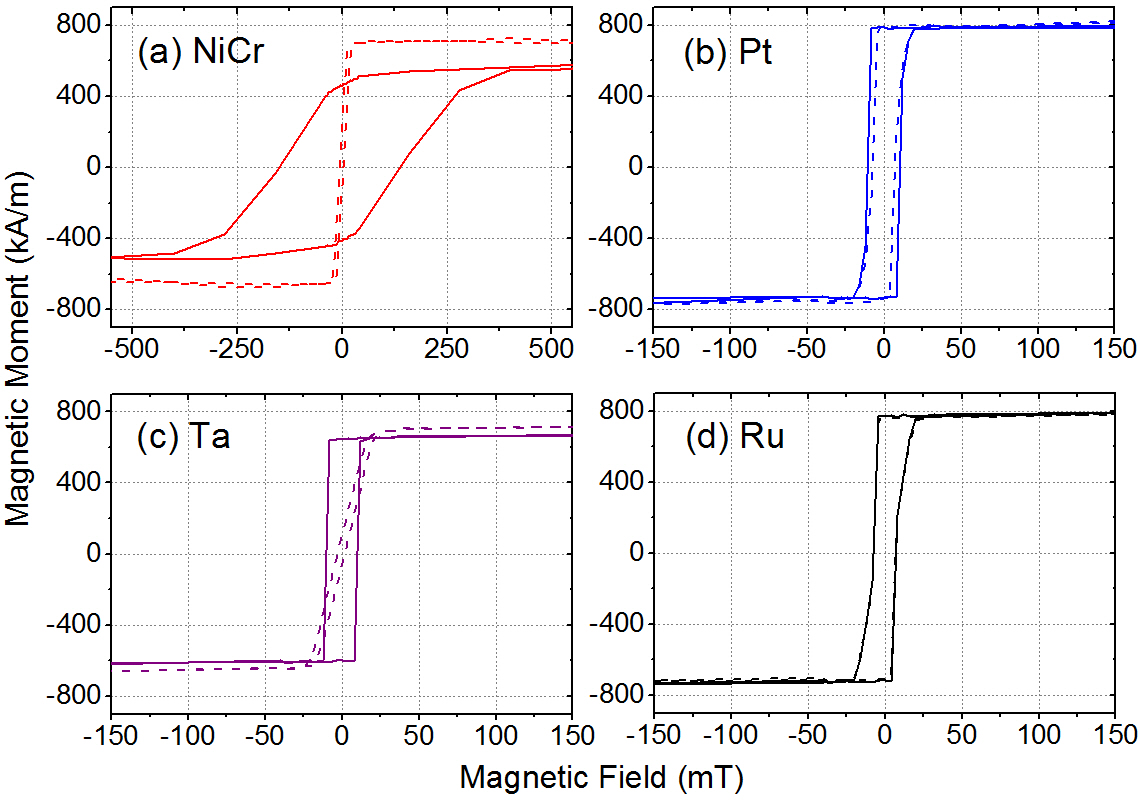}
\caption{Perpendicular magnetization loops of [Co/Ni] on different seed layers as-deposited (dashed lines) and after 300$^{\circ}$C annealing (solid lines) by VSM: (a) NiCr, (b) Pt, (c) Ta and (d) Ru. Note that the \textit{x}-axis scale in (a) is different from the others.
}
\label{F4}
\end{figure}

\begin{figure}
\centering
\includegraphics[scale=0.38]{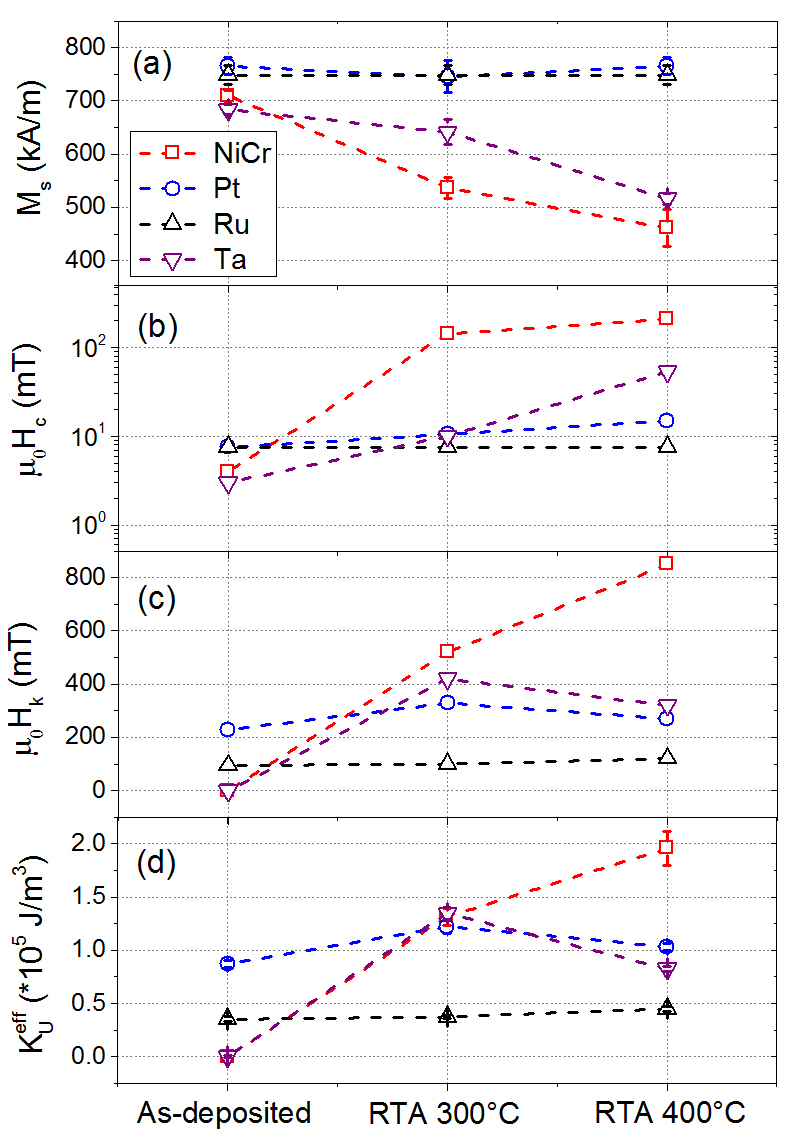}
\caption{Magnetic properties of [Co/Ni] on different seed layers under different annealing conditions:  (a) $M_s$, (b) $\mu_0H_c$, (c) $\mu_0H_k^{\mathrm{eff}}$ and (d) $K_U^{\mathrm{eff}}$.}
\label{F5}
\end{figure}

The presence of PMA in [Co/Ni] before and after annealing is firstly checked by VSM (Fig.\ref{F4}). Despite the presence of \textit{fcc}(111) peaks on all seeds, no PMA was observed in the as-deposited films on NiCr and Ta seeds. In contrast, PMA occurs in as-deposited [Co/Ni] on Pt and Ru seed. After 300$^{\circ}$C annealing, PMA appears in [Co/Ni] on NiCr and Ta seed (see Fig.4(a) and (c), respectively). Simultaneously, an $M_s$ loss is observed. Note that for the [Co/Ni] on NiCr, the $M_s$ loss is large and the hysteresis loop becomes bow-tie like with coercivity ($\mu_0H_c$) increase (Fig.\ref{F4}(a)). The large $\mu_0H_c$ enables [Co/Ni] on NiCr seed to function as hard layer in MTJ stacks\cite{Kar2014,Swerts2015,Tomczak2016}. Fig.\ref{F5}(a) and (b) summarizes $M_s$ and $\mu_0H_c$ of the [Co/Ni] on different seed layers for various annealing conditions. The $M_s$ and $\mu_0H_c$ of [Co/Ni] on NiCr and Ta seed decrease and increase, respectively, further after 400$^{\circ}$C annealing. For the [Co/Ni] Pt and Ru seed, there is no change in $M_s$ and $\mu_0H_c$(see Fig.\ref{F4}(b) and (d), respectively), even after 400$^{\circ}$C annealing, indicating a good thermal tolerance.

Fig.5(c) and (d) show the effective perpendicular anisotropy field ($\mu_0H_k^{\mathrm{eff}}$) values and the calculated  $K_U^{\mathrm{eff}}$ of [Co/Ni] on each seed for different annealing conditions. As deposited, the $\mu_0H_k^{\mathrm{eff}}$ of [Co/Ni] on NiCr and Ta seed is 0, meaning that they have IMA, as shown in Fig.\ref{F4}, while the highest $\mu_0H_k^{\mathrm{eff}}$ is found on Pt seed as expected from the small lattice mismatch and the crystalline nature of Pt buffers, i.e. \textit{fcc}(111). After annealing at 300$^{\circ}$C, $\mu_0H_k^{\mathrm{eff}}$ significantly increases in [Co/Ni] on all seeds, except when the [Co/Ni] is grown on Ru. On Ru, $\mu_0H_k^{\mathrm{eff}}$ is the lowest. After 400$^{\circ}$C annealing, PMA is maintained in all samples, though $\mu_0H_k^{\mathrm{eff}}$ of [Co/Ni] on Pt and Ta seed slightly decrease. For the NiCr seed, $\mu_0H_k^{\mathrm{eff}}$ even increases further and becomes more than 2 times larger than the values found in the other samples. The large $\mu_0H_k^{\mathrm{eff}}$ and  $K_U^{\mathrm{eff}}$ of [Co/Ni] on NiCr, especially after 400$^{\circ}$C, will be analyzed in the following.

\subsection{Correlation between structural and magnetic properties of [Co/Ni]}
Commonly, highly \textit{fcc}(111) textured [Co/Ni] films result in large PMA. In our case, we have observed some anomalous behaviors. The as-deposited film on NiCr and Ta did not show PMA. The PMA occured and increased significantly after annealing, while $M_s$ loss was observed. On the other hand, only limited PMA increase is observed on Pt seed after annealing, despite the large increase in diffraction peak intensity shown in Fig.\ref{F1}(b). In short, the improvement of [Co/Ni] film quality leads to limited increase in PMA for [Co/Ni] on Pt and Ru seed, yet there is a huge increase in PMA for [Co/Ni] on NiCr and Ta seed. In the further, we will discuss the mechanisms responsible for the observed trends. 

\subsubsection{Impact of strain on the static magnetic properties}
As shown in Section III.A, the \textit{fcc}(111) peak position of [Co/Ni] on the various seed layers differs from each other, indicating the existence of strain induced by the seed layer. Because of the magneto-elastic effect, strain-induced magnetic anisotropy ($K_s$) can be an important contribution of the total PMA. $K_s$ is calculated as\cite{Sander1999,Gopman2016}
\[K_s = \frac{18B_2^{\mathrm{Co}}+30B_2^{\mathrm{Ni}}}{48}\cdot(\varepsilon_0-\varepsilon_3).\] In this equation, $B_2^{\mathrm{Co}}=-29$MJ/m$^3$ and $B_2^{\mathrm{Ni}}=+10$MJ/m$^3$ reflect the \textit{fcc}(111) cubic magneto-elastic coupling coefficients of bulk Co and Ni, respectively\cite{Sander1999}. Possible thin film effects on the coefficients are beyond the scope of the paper. $\varepsilon_3$ is the out-of-plane strain, which can be derived from the shift of \textit{fcc}(111) peak in XRD, with $d_{111} = 2.054\mathrm{\AA}$ as the reference\cite{davis2000nickel}. And $\varepsilon_0=-\varepsilon_3/\nu$, where $\nu$ is Poisson ratio. $\nu$ is calculated as weight-averaged values of bulk Co and Ni\cite{cardarelli2008materials}. 

\begin{figure}
\centering
\includegraphics[scale=0.26]{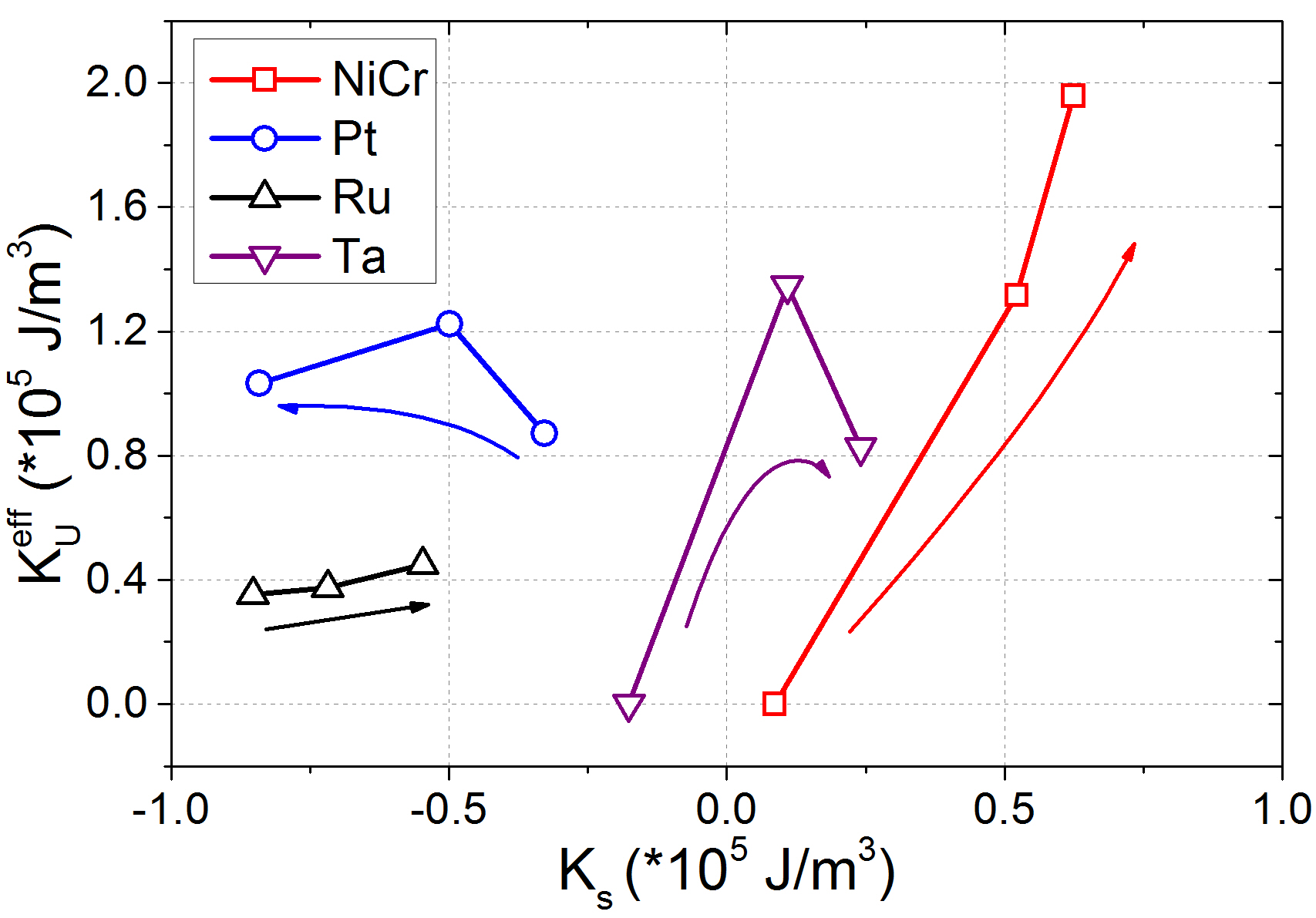}
\caption{The relation between effective perpendicular anisotropy energy ($K_U^{\mathrm{eff}}$) and strain-induced anisotropy energy ($K_s$). The arrows indicate the temperature of annealing conditions: as-deposited, 300$^{\circ}$C, and 400$^{\circ}$C.}
\label{F6}
\end{figure}

Fig.\ref{F6} summarizes the strain-induced PMA before and after annealing. It is clear that the strain from Pt and Ru seed result always in a negative contribution to the PMA of [Co/Ni], which may in both cases counteract with the increase in PMA from improved film quality after annealing (see the narrower peak of [Co/Ni] with larger intensity after annealing in Fig.\ref{F1} and decreased FWHM in Fig.\ref{F2}) and results in little net $K_U^{\mathrm{eff}}$ improvement (see Fig.\ref{F5}(d)). Similarly observed in Fig.\ref{F5}(d), as-deposited [Co/Ni] on Ta seed shows low $K_U^{\mathrm{eff}}$ due to the negative strain-induced PMA, even though it has the required texture (see Fig.\ref{F1}). After annealing, strain-induced PMA contributes positively to total $K_U^{\mathrm{eff}}$ of [Co/Ni] on Ta seed. For [Co/Ni] on NiCr seed, the strain after annealing promotes the increase in total $K_U^{\mathrm{eff}}$. From this discussion, it is clear that the seed layer providing in-plane tensile strain to [Co/Ni] is desired for PMA increase.

However, only the strain contribution cannot explain the large PMA that is observed after annealing on NiCr. Moreover, as shown in Fig.\ref{F5}(b), there is also a large increase in $\mu_0H_c$ for [Co/Ni] on NiCr and Ta samples, while their $M_s$ reduce dramatically, which cannot be explained by the previous strain-induced PMA change. 

\subsubsection{Impact of diffusion on the static magnetic properties}
\begin{figure*}
\centering
\includegraphics[scale=0.43]{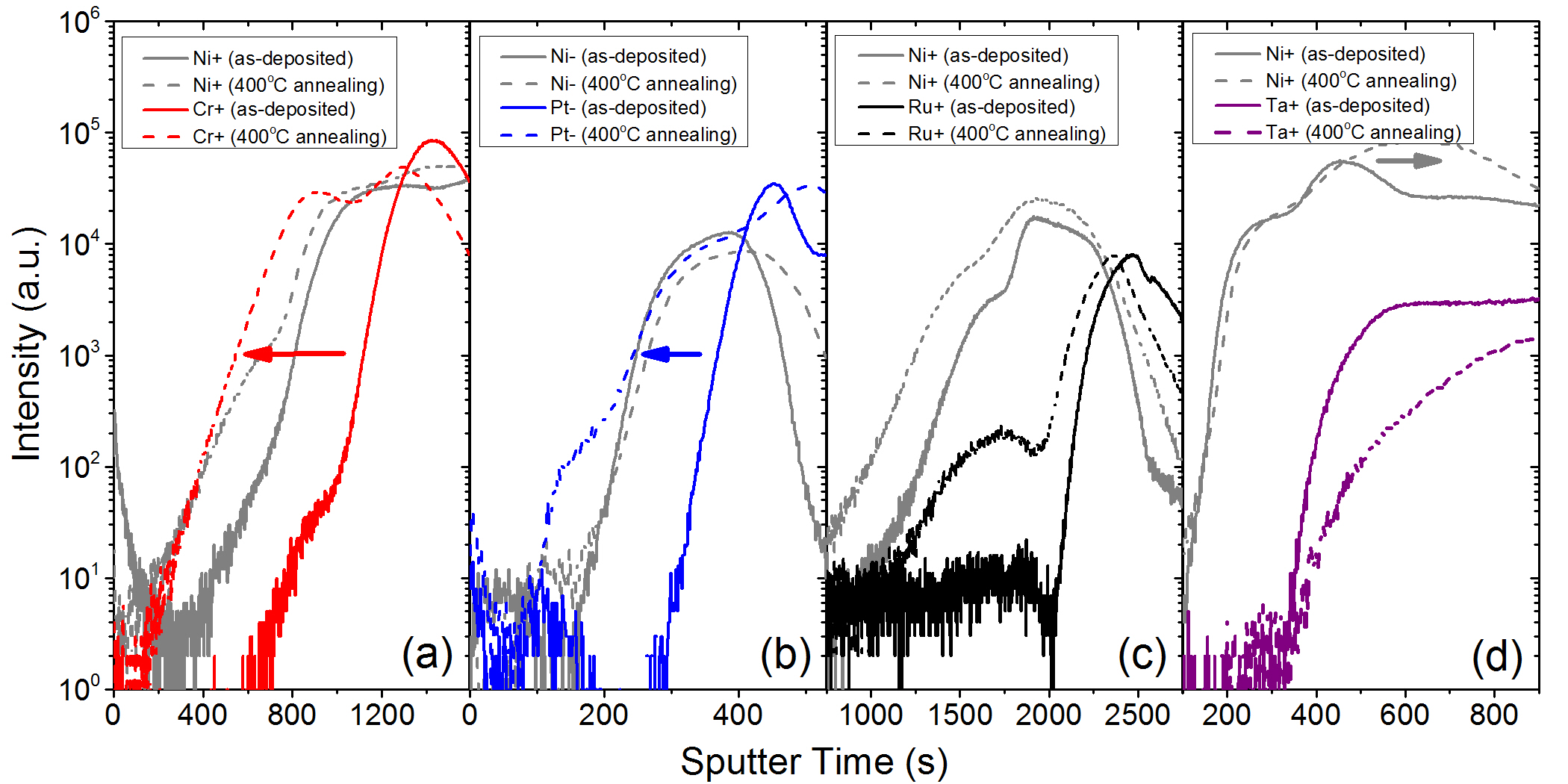}
\caption{ToF-SIMS depth profiles of the seed layer elements in [Co/Ni] stacks: (a) Cr, (b) Pt, (c) Ru and (d) Ta. Arrows in the figures indicate the shift of signal of corresponding elements before and after annealing.}
\label{F7}
\end{figure*}

We have performed an advanced compositional analysis of the [Co/Ni] after annealing. Fig.\ref{F7} shows the ToF-SIMS depth profiles of Cr, Pt, Ru and Ta as-deposited and after 400$^{\circ}$C annealing. The Ni signal is provided to indicate position of the [Co/Ni] multilayers. In the case of the NiCr seed, Cr is found throughout the whole layer of [Co/Ni] after 400$^{\circ}$C annealing, since its signal appears at the same depth (sputter time) as Ni. That means Cr diffuses heavily in the [Co/Ni]. The same phenomenon is observed for Pt seed, but the intensity of the signal from diffused Pt is low when compared to the Pt signal in the seed layer part, indicating that the diffusion amount of Pt is limited. For Ru shown in Fig.\ref{F7}(c), the interface between the Ru seed and [Co/Ni] remains sharp after annealing. On the contrary, the less steep increase in Ta signal suggest intermixing between [Co/Ni] and Ta after annealing at the interface, but does not suggest Ta diffusion in the bulk of the [Co/Ni] films. 

\begin{figure}
\centering
\includegraphics[scale=0.36]{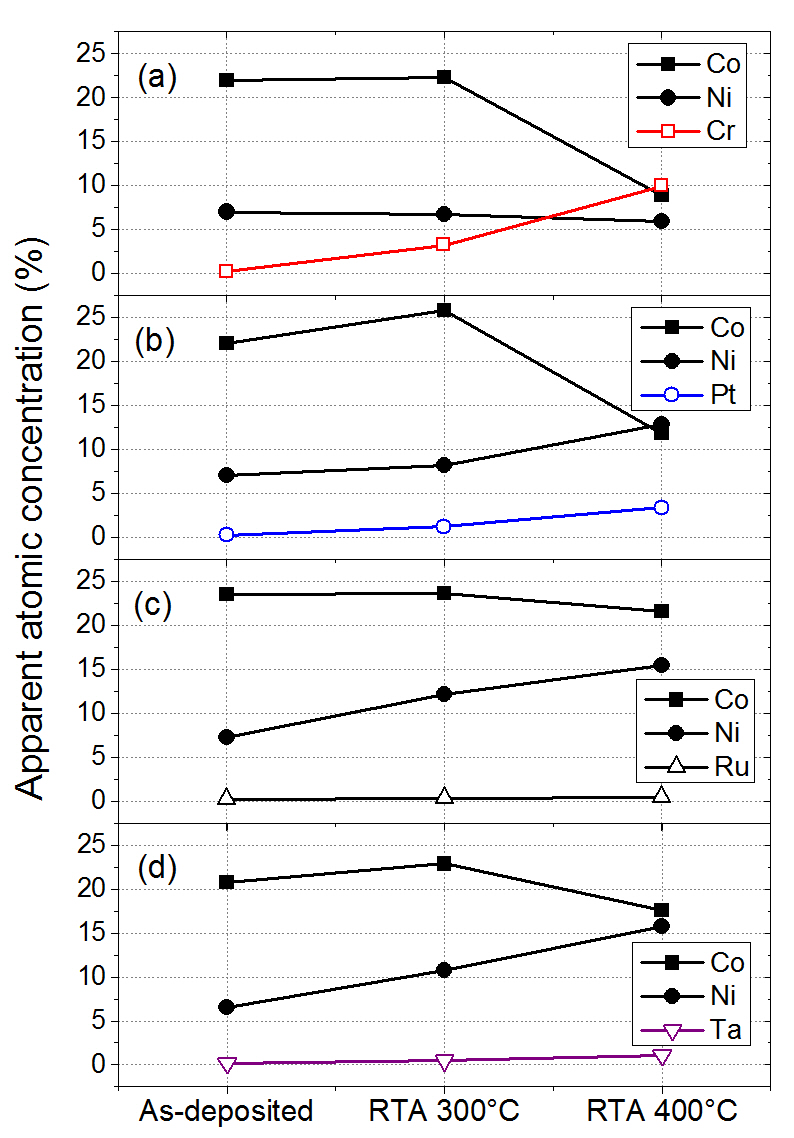}
\caption{Apparent atomic concentration of Co, Ni and corresponding seed layer element in the stack before and after annealing: (a) Cr, (b) Pt, (c) Ru and (d) Ta. The values are integrated over a few nm (electron mean free path), taking into account a exponentially decreasing sensitivity of the XPS signal with the depth. Other elements like C and O, which have no influence on the property change of [Co/Ni], are omitted in the figure.}
\label{F8}
\end{figure}

To quantitatively study the diffusion of the seed in the [Co/Ni], XPS measurements are conducted and the apparent atomic concentration of Co, Ni and seed layer element in each sample with different annealing conditions are shown in Fig.\ref{F8}. Note that the higher apparent concentration of Co when compared to Ni for the as-deposited sample is due to the surface sensitivity of the XPS technique as explained in the figure caption. It is clear that Cr diffuses the most among the four seed layers. The presence of Cr in [Co/Ni] may also lead to the shift of the [Co/Ni] peak towards the NiCr peak in the XRD pattern shown in Fig.\ref{F1}(a), probably resulting in the formation of Co-Ni-Cr alloy. Furthermore, the observed change in magnetic properties can likely be attributed to the formation of the Co-Ni-Cr alloy\cite{Ishikawa1986,Tokushige1990}. Indeed, the uniform diffusion of Cr was reported to cause $\mu_0H_c$ increase in Co-Ni film with in-plane anisotropy\cite{Hasegawa1989}. And Cr can be coupled antiferromagnetically with its Co and Ni hosts causing the $M_s$ drop\cite{Iwasaki1978} and higher PMA. Less diffusion is observed for the Pt seed, in agreement with the peak shift toward Pt seed (Fig.1(b)) and for the lower increase in PMA as well, since the [Co/Pt] system, alloys or multilayers, is a well-known PMA system and so a change in PMA due to Pt atoms in the [Co/Ni] matrix is not necessarily detrimental\cite{Chen2015}. On the contrary, no significant diffusion of Ru into [Co/Ni] layers has been observed in Fig.\ref{F8}(c), so no impact on PMA is to be expected. Finally, a large increase in PMA has been observed after annealing whereas the diffusion is limited for Ta seed. Possibly, the improvement of crystal structure and the formation of a Co-Ni-Ta alloy at the interface of Ta seed and [Co/Ni] part happens at higher temperature, which gives rise to PMA and $M_s$ loss\cite{Khan1990,Gupta2001, Kil2015}. 

Apart from the $M_s$ and PMA change, the diffusion of the seed material into the [Co/Ni] might also impact the magneto-elastic coefficients to be taken into account when calculating the strain-induced PMA (see Section III.C.1). This impact is however not straightforward and requires further study.

\subsubsection{Impact of diffusion on the dynamic magnetic properties}
\begin{figure}
\centering
\includegraphics[scale=0.1]{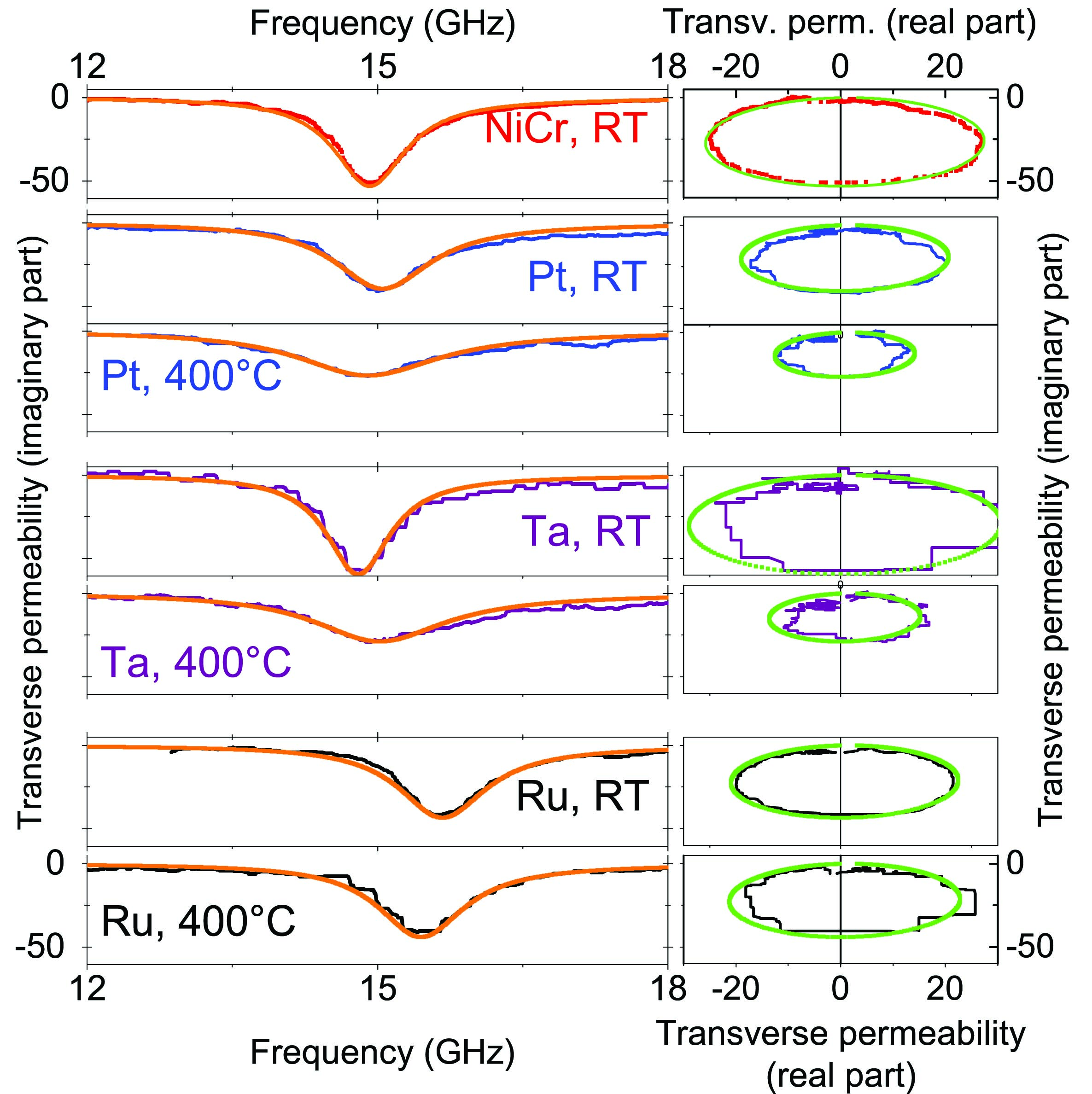}
\caption{FMR linewidth for out-of-plane field conditions leading to resonances near 15 GHz for as-deposited and annealed samples. Left panels: imaginary part of the permeability spectra (symbols) and macrospin fits thereof (orange lines). Right panels: polar plots of the imaginary part versus real part of the transverse permeability. The green symbols are the macrospin fits\cite{Devolder2013}, i.e. there are essentially circles of diameters proportional to $M_s/\alpha$.}
\label{F9}
\end{figure}

Earlier studies reported on dopants that increase the damping when incorporated into a ferromagnetic film\cite{Rantschler2007,Devolder2013}. In our case, it is natural to expect that an impact from the diffused seed layer element will be exerted on the dynamic magnetic properties of [Co/Ni]. Therefore, the Gilbert damping constant ($\alpha$) of each sample with different annealing conditions was derived from VNA-FMR for study. Fig.\ref{F9} compares the permeabilities of the [Co/Ni] films as-deposited and after 400$^{\circ}$C annealing when the FMR frequency is set to 15 GHz by a proper choice of the applied field. The broadening of the linewidths after annealing reflects the increase in damping for all cases, with the noticeable exception of Ru. The NiCr resonance was too broadened to be resolved after 400$^{\circ}$C annealing, reflecting a very high damping or a very large inhomogenity in the magnetic properties. Linear fits of the FMR linewidth versus FMR frequency were conducted (not shown) to extract the damping parameters, which are listed in Table \ref{T1}. It should be noticed that in our VNA-FMR measurement, the two-magnon contributions to the linewidth and hence its impact on damping derivation can be excluded due to the perpendicular geometry in the measurement\cite{Shaw2014}. And the contribution of spin-pumping within the seed layer to the linewidth in our cases is always expected to be within the error bar\cite{Zwierzycki2005}. The lowest damping in the as-deposited [Co/Ni] films were obtained on Ta and NiCr seed, i.e. in the in-plane magnetized samples. The highest damping was found on Pt seed, a fact that is generally interpreted as arising from the large spin-orbit coupling of Pt. And the damping values increase with post annealing in [Co/Ni] on all seed layers except Ru. Though the damping of [Co/Ni] is larger than CoFeB/MgO\cite{Liu2011a}, it is equal to or smaller than [Co/Pt] and [Co/Pd]\cite{Kato2012,Ishikawa2016}, which makes it of interest to use as free layer material with high thermal stability in high density STT-MRAM applications\cite{Liu2016}. 

\begin{table}
\caption{\label{T1} $\alpha$ provided by VNA-FMR for as-deposited and post-annealed [Co/Ni] on different seed layers. }
\begin{ruledtabular}
\begin{tabular}{ccccc}
 & Seed & As-depo & RTA 300$^{\circ}$C & RTA 400$^{\circ}$C\\
\hline 
\rule[1ex]{0pt}{2.5ex}                  
\multirow{4}{*}{$\alpha$ ($\pm$10\%)} & NiCr & 0.020 & 0.040  & $\gg$0.040 \\
                  & Pt & 0.026  & 0.030 & 0.034 \\
                  & Ru & 0.022 & 0.020 & 0.018 \\
                  & Ta & 0.017 & 0.024 & 0.026 
\end{tabular}
\end{ruledtabular}
\end{table}

\begin{figure}
\centering
\includegraphics[scale=0.28]{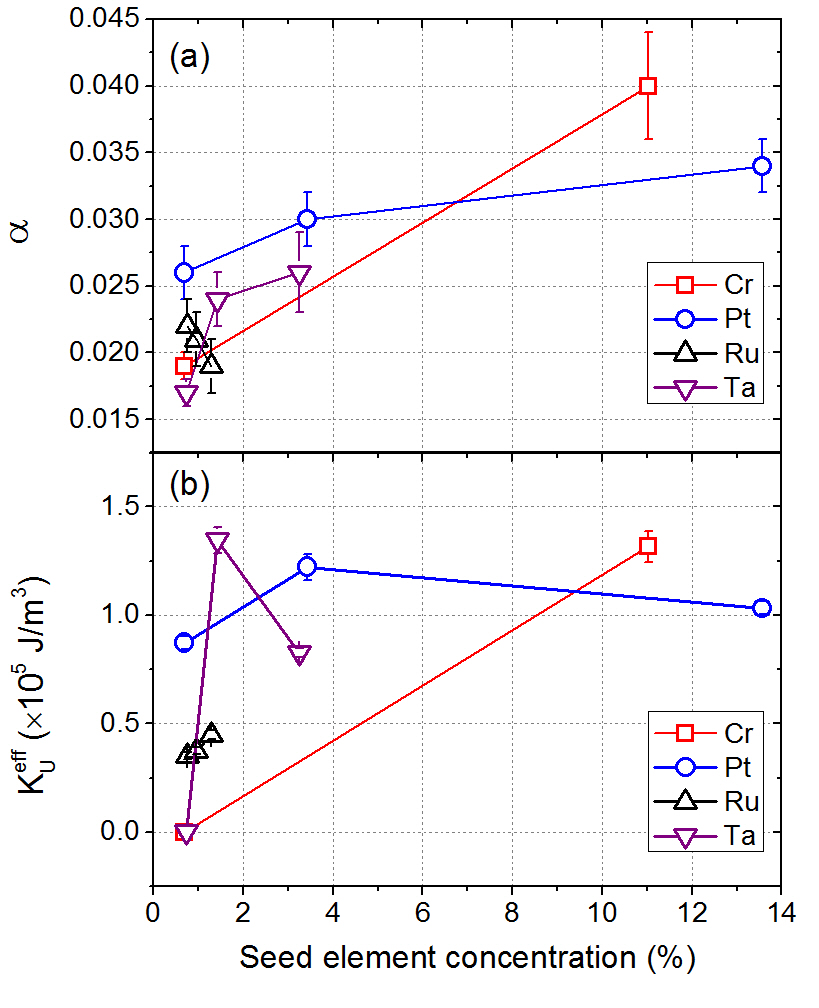}
\caption{(a) $\alpha$ and (b) $K_{U}^{\mathrm{eff}}$ as a function of seed element concentration in [Co/Ni]. For each series, the annealing condition from left to right is as-deposited, RTA 300$^{\circ}$C and RTA 400$^{\circ}$C. The 400$^{\circ}$C data of NiCr are not shown (Cr concentration $\sim$60\%).}
\label{F10}
\end{figure}

Fig.\ref{F10}(a) and (b) plot the correlations between the dopant concentration and the $\alpha$ and $K_U^{\mathrm{eff}}$ values, respectively. Here the seed element concentration in the [Co/Ni] multilayer is reflected by the ratio of the apparent seed concentration (at.\%) divided by the sum of the apparent Co and Ni concentrations (at.\%) for each sample and annealing condition. There is a clear correlation between the damping and the concentration of the seed element in the [Co/Ni] system. The absence of evolution of the damping upon annealing for [Co/Ni] on Ru can thus be explained by the non-diffusive character of the Ru seed. In the case of the NiCr seed, the damping is the largest after annealing. A similar trend was observed before on Au seeds and attributed to the formation of superparamagnetic islands in thin [Co/Ni] multilayers\cite{Kurt2010}. However, since we observe for the same [Co/Ni] system no $M_s$ loss and no $\mu_0H_c$ increase on Ru and Pt, and since the PMA on NiCr after anneal is the highest, the formation of super paramagnetic islands cannot be the reason to explain the magnetic behavior. It is more likely that the increased damping comes from the formation of a textured Co-Ni-Cr alloy with high PMA and high damping, as mentioned in Section III.C.2. Finally, in the case of Pt and Ta seeds, the damping on the Pt seed is higher than on the Ta seed for the same concentration. That is attributed to the spin-orbit coupling for Pt dopants being substantially higher than for Ta. In conclusion, it is clear that the seed diffusion cannot be ignored when studying the impact of annealing on the structural and magnetic properties of [Co/Ni].   

\section{conclusions}
\begin{figure}
\centering
\includegraphics[scale=0.58]{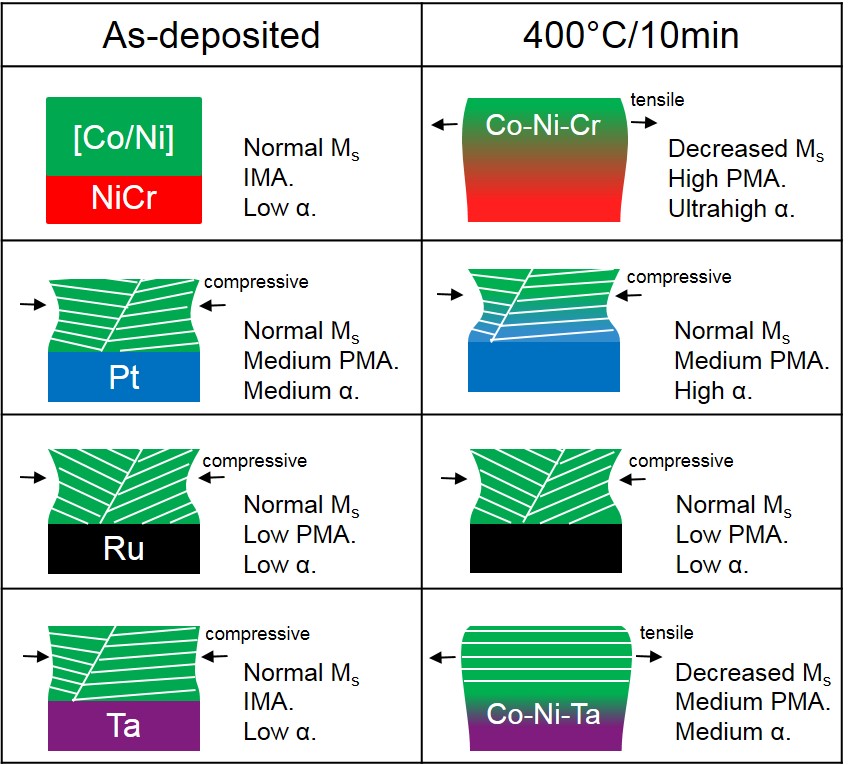}
\caption{Schematic summary of the impact of annealing impact on various seeds on the structural and magnetic properties of [Co/Ni].}
\label{F11}
\end{figure}

In summary, the structural and magnetic properties of [Co/Ni] on different seed layers are investigated. [Co/Ni] on Pt and Ru seed show \textit{fcc}(111) texture after deposition and have PMA. IMA is observed for [Co/Ni] on NiCr and Ta seed, and PMA appears and increases after post-annealing. Further annealing  improves the texture and hence increases PMA. Meanwhile, the shift of [Co/Ni] diffraction peaks in XRD curves indicates the presence of strain in the [Co/Ni] film, which can also influence PMA. The strain-induced PMA may have a positive effect on the total PMA (NiCr and Ta seed), or a negative impact (Pt and Ru seed). A dramatic reduction of $M_s$ and large increase in $\mu_0H_c$ of [Co/Ni] on NiCr and Ta seed after annealing is observed. The damping property of [Co/Ni] on different seed layers evolves similarly as PMA with post-annealing. The explanation for these phenomena is the diffusion property of seed layer materials. High PMA and very large damping is obtained on NiCr seed because of the dramatic diffusion of Cr and the formation of Co-Ni-Cr alloy. Though non-diffusive Ru seed results in low damping, low PMA as-deposited and after annealing is obtained. Pt seed can provide good PMA, but its large spin-orbit coupling exerted from the interface with [Co/Ni] and diffused part increases damping, especially after annealing. PMA as high as on Pt has been observed for Ta seed with lower damping due to its smaller spin-orbit coupling. Fig.\ref{F11} summarizes schematically the impact of seed layer and annealing on the structural and magnetic properties of [Co/Ni] on different seed layers.

Finally, by selecting the seed and post-annealing temperature, the [Co/Ni] can be tuned in a broad range from low damping to high damping while maintaining PMA after annealing until 400$^{\circ}$C.  As such, the [Co/Ni] multilayer system is envisaged for various applications in spintronics, such as highly damped fixed layers, or low damping free layers in high density magnetic memory or domain wall motion mediated spin logic applications.\\

This work is supported by IMEC's Industrial Affiliation Program on STT-MRAM devices.

\bibliographystyle{apsrev4-1}
%

\end{document}